\documentclass[]{aa}
\usepackage{natbib,graphicx}

\def\etal{{\it et al.\ }}

\def\kms{\ {\rm km\,s^{-1}}}

\begin{document}

\title{The cluster of galaxies Abell 376.
\thanks{based on observations made Haute-Provence and Pic du Midi
Observatories (France)}}
\author{D.~Proust\inst{1} \and H.V. Capelato\inst{2} \and G. Hickel\inst{2}
\and L.~Sodr\'e Jr.\inst{3} \and G.B.~Lima~Neto\inst{3}
\and H.~Cuevas\inst{3,4}}

\offprints{D.~Proust} 

\institute{%
Observatoire de Paris-Meudon, GEPI, F92195 MEUDON, France 
\and Divis\~ao de Astrof\'{\i}sica, INPE/MCT, 12227-010, 
     S\~ao Jos\'e dos Campos/S.P., Brazil
\and Instituto de Astronomia, Geof\'{\i}sica e Ci\^encias Atmos., 
     Universidade de S\~ao Paulo (IAG/USP), 01065-970 S\~ao Paulo/S.P., Brazil 
\and Universidad La Serena, Casilla 544, La Serena, Chile.}

\date{Received date; accepted date}

\abstract{%
We present a dynamical analysis of the galaxy cluster Abell~376 based on a set
of 73~velocities, most of them measured at Pic du Midi and Haute-Provence
observatories and completed with data from the literature. Data on individual
galaxies are presented and the accuracy of the determined velocities is
discussed as well as some properties of the cluster. We obtained an improved 
mean redshift value $z = 0.0478^{+0.005}_{-0.006}$ and velocity dispersion 
$\sigma = 852^{+120}_{-76}\kms$.
Our analysis indicates that inside a radius of $\sim 900 h_{70}^{-1}\,$kpc
($\sim 15\,$arcmin) the cluster is well relaxed without any remarkable feature
and the X-ray emission traces fairly well the galaxy distribution. A possible
substructure is seen at 20 arcmin from the centre towards the Southwest
direction, but is not confirmed by the velocity field. This SW clump is,
however, kinematically bound to the main structure of Abell~376. A dense
condensation of galaxies is detected at 46 arcmin (projected distance $2.6\,
h_{70}^{-1}\,$Mpc) from the centre towards the Northwest and analysis of the
apparent luminosity distribution of its galaxies suggests that this clump is
part of the large scale structure of Abell~376.
X-ray spectroscopic analysis of ASCA data resulted in a temperature $kT = 4.3
\pm 0.4\,$keV and metal abundance $Z = 0.32 \pm 0.08 Z_{\odot}$. The velocity 
dispersion corresponding to this temperature using the $T_{X}$--$\sigma$ 
scaling relation is in agreement with the measured galaxies velocities.
\keywords{galaxies: distances and redshifts -- galaxies: cluster: general --
galaxies: clusters: Abell 376 -- clusters: X-rays -- clusters: subclustering}
}

\maketitle

\section{Introduction} \label{Introduction}

Redshift surveys in clusters of galaxies are needed to study their dynamical
and evolutionary states, estimating parameters such as the mass, shape and
distortion of the velocity field, presence of substructures or projected
foreground/background galaxies and groups, strength of dynamical friction and
two-body processes and, in general, the present stage of their dynamical
evolution. This information is useful to test theoretical scenarios of galaxy
formation and also of large structure formation and evolution.

In clusters, the mean velocity is a key factor in deriving distances, allowing
the study of matter distribution over very large scales. Within clusters the
analysis of the velocity field can lead to an estimate of the virial mass,
constraining models of the dark matter content. Galaxy velocity measurements
provide information complementary to that obtained through X-ray observations
of clusters. Both form basic pieces of information for the understanding of
clusters. However, a discrepancy between these estimators are often found
(e.g. Mushotzky \etal 1995; Girardi \etal 1998; Wu \etal 1999). Virial mass
estimates rely on the assumption of dynamical equilibrium. X-ray mass
estimates also depend on the dynamical equilibrium hypothesis and on the still
not well-constrained intra-cluster gas temperature gradient (e.g., De Grandi
\etal 2002). Finally, mass estimates based on gravitational strong lensing are
considered more reliable than the others (e.g., Mellier 1999) because they are
completely independent of the dynamical status of the cluster. The drawback is
that strong lensing can only probe the central region of clusters. The
discrepancies with other methods may be due to non-equilibrium effects in the
central region of the clusters (Allen, 1998).

An important source of departure from equilibrium (that may affect mass
estimates) are substructures. Their very existence supports the current view
that clusters grow hierarchically by accreting nearby groups and galaxies.
Even the frequency and degree of clumpiness in the central regions of the
clusters depends on the cosmology (e.g. Richstone \etal 1992). In many cases,
substructures are loosely bound and can survive only a few crossing times in
the hostile environment of rich clusters. However, they seem to be very common
in present-day clusters. A recent estimate by Kolokotronis \etal (2000)
indicates that at least 45\% of rich clusters present optical substructures.
Substructures are detected in both optical and X-ray images in 23\% of the
clusters. This last number may then be considered a lower limit on the
fraction of real substructures in clusters and implies that one in four
clusters may be out of equilibrium due to the presence of a substructure. The
dynamical status of individual clusters should therefore be examined in detail
before being used in other studies.

In this paper, we present a study of the dynamical status of the cluster
Abell~376, from the analysis of the positions and velocities of its galaxies,
as well as from the intra-cluster gas X-ray emission. Abell~376 has a I-II
morphology in the Bautz-Morgan classification and a richness class R~=~0
extending on 151 arcmin in diameter (Abell \etal 1989). It is a cD or D-galaxy
dominated cluster; its brightest cluster member (BCG) was recently imaged by
the HTS WFPC2 (Laine et al. 2003). It shows a smooth brightness profile with a
central core, although it is not well fitted by the so-called Nuker-law (Lauer
et al. 1995). Abell~376 was observed in X-rays by several satellites. It has a
moderate cooling flow $M_{cool} = 42^{+42}_{-14} M_{\odot}$yr$^{-1}$ (White
\etal 1997).

A search in the NED database\footnote{The NASA/IPAC Extragalactic Database
(NED) is operated by the Jet Propulsion Laboratory, California Institute of
Technology, under contract with the National Aeronautics and Space
Administration.} indicates that only few radial velocities were known in the
field of the cluster. A list of 13~velocities is published in Capelato \etal
(1991) obtained at Haute-Provence observatory with the multiobject
spectrograph SILFID for which we noticed a few years after that except the
central E/D number~71 galaxy, there was misidentification in the
correspondence object/fiber/spectrum, so that the velocities cannot be used
and should be removed from any database. For this reason, we decided to
reobserve some of the previous galaxies and not take into account our previous
velocities. Here we examine some properties of the cluster Abell~376, using a
set of 73~new and already published velocities. The observations of radial
velocities reported here are part of a program to study the dynamical
structure of clusters of galaxies, started some years ago and with several
results already published (see e.g. Proust \etal 1992, 1995, 2000; Capelato
\etal 1991, Sodr\'e \etal 2001).

This paper is organized as follows. We present in Sect.~2 the details of the
observations and data reduction. In Sect.~3 and 4 we discuss the distribution
and the velocity analysis of the cluster galaxies respectively. In Sect.~5 we
analyze the X-ray emission and gas distribution and in Sect.~6 the spectral
and morphological classification of the cluster galaxies. In Sect.~7 we
present mass estimates for the central region of the cluster, derived from
optical and X-ray observations and the dynamical status of Abell~376. Finally,
in Sect.~8 we summarize our conclusions. We adopt here, whenever necessary,
$H_{0}= 70\, h_{70}\kms$Mpc$^{-1}$, $\Omega_{0} = 0.3$ and $\Omega_{\Lambda} =
0.7.$

\section{Observations and Data Reductions}\label{Observations and Data Reductions}

The new velocities presented in this paper have been obtained with the 1.93m
telescope at Haute-Provence Observatory and with the 2.0m telescope at Pic du
Midi (France).

Observations with the 1.93m Haute-Provence Observatory telescope were carried
out in November 1997 and 1998. We used the CARELEC spectrograph at the
Cassegrain focus, equipped with a 150~lines/mm grating blazed at 5000~{\AA}
and coupled to an EEV CCD detector 2048x1024 pixels with pixel size of
13.5~$\mu$m. A dispersion of 260~{\AA}/mm was used, providing spectral
coverage from 3600 to 7300~{\AA}. Wavelength calibration was done using
exposures of He-Ne lamps.

We completed the observations during an observing run at the 2.0m Bernard Lyot
telescope at Pic du Midi observatory in January 1997 with the ISARD
spectrograph in its long-slit mode with a dispersion of 233~\AA/mm with the
TEK chip of 1024x1024 pixels of 25$\mu$m corresponding to 5.8~\AA/pixel.
Typically two exposures of 2700s each were taken for fields across the
cluster. Wavelength calibration was done using Hg-Ne lamps before and after
each exposure.

The data reduction was carried out with IRAF\footnote{IRAF is distributed by
the National Optical Astronomy Observatories, which are operated by the
Association of Universities for Research in Astronomy, Inc., under cooperative
agreement with the National Science Foundation.} using the LONG SLIT package.
The spectra were rebinned with a scale of 1~{\AA}/bin equally spaced in log
wavelength. Radial velocities have been determined using the cross-correlation
technique (Tonry and Davis 1979) implemented in the RVSAO package (Kurtz \etal
1991, Mink \etal 1995) with radial velocity standards obtained from
observations of late-type stars and previously well-studied galaxies.

A total of 71 velocities has been obtained from our observations.
Table~1\footnote{Table~1 is also available in electronic form at the CDS via
anonymous ftp 130.79.128.5} lists positions and heliocentric velocities for 73
individual galaxies in the cluster including data from Wegner \etal (1991) and
from Postman an Lauer (1995), with the following columns:

\makeatletter\if@referee\renewcommand\baselinestretch{1.0}\fi\makeatother
\begin{table*}[htb]
\caption[]{Heliocentric redshifts, position and morphological type for
galaxies of Abell~376.}
\begin{tabular}{llllllll}
\hline
\hline
Galaxy & Name & R.A. & Decl. & Type & Hel. Vel. & R & N \\
 id.   &     & (2000) & (2000) &    & $V {\pm {\Delta}V}$  & & \\
\hline
  2 &             & 02 46 24.7 & +36 28 24 &    E   & 39170  78 &  3.92 & o  \\
  3 &             & 02 46 24.3 & +36 28 21 &    E   & 39711 105 &  5.60 & o  \\
  5 &             & 02 45 30.1 & +36 28 33 &   SB   & 14222  62 &  6.71 & o  \\
  6 &             & 02 47 29.6 & +36 35 29 &    E   & 39809  76 &  4.13 & o1 \\
  7 &             & 02 47 25.8 & +36 36 23 &  S0/E  & 32510  69 &  4.76 & o  \\
  8 &             & 02 47 24.6 & +36 34 49 &   S0   & 14452  24 &  7.31 & o  \\
  9 &             & 02 46 48.5 & +36 32 47 &   S0   & 14906  31 &  4.63 & o  \\
 11 &             & 02 46 37.2 & +36 33 25 &    E   & 14856  70 &  6.02 & o  \\
 17 &             & 02 44 43.9 & +36 35 46 &  E/S0  & ~~~---    &  5.56 & o  \\
 18 &             & 02 47 35.2 & +36 40 55 &    S   & 13660  68 &  6.69 & o  \\
 20 &             & 02 46 02.5 & +36 41 27 &   S0   & 14314  45 &  6.65 & o  \\
 22 &             & 02 45 51.8 & +36 39 29 &   S0   & 14828  31 & 12.44 & o  \\
 24 &             & 02 45 21.9 & +36 42 58 &   S0   & 13419 130 &  2.85 & o  \\
 25 & GIN 139     & 02 45 13.7 & +36 41 41 &  S0/E  & 15010  33 &  5.92 & o  \\
    &             &            &           &        & 15002  21 &       & l1 \\
 26 & GIN 140     & 02 45 12.4 & +36 42 44 &   S0   & 13960  18 &       & l1 \\
 27 &             & 02 45 10.3 & +36 41 39 &  Sa/0  & 13437  41 &  8.79 & o  \\
 28 & GIN 141     & 02 45 03.9 & +36 42 36 &    E   & 14844  38 & 11.73 & o  \\
    &             &            &           &        & 14822  24 &       & l1 \\
 31 &             & 02 47 46.6 & +36 43 55 &  S0/a  & 14286  60 &  7.85 & o  \\
 32 &             & 02 47 21.7 & +36 43 40 &    S   &  5130  79 &  3.56 & o  \\
 33 &             & 02 47 04.2 & +36 45 35 &   S0   & 13655  49 &  6.86 & o  \\
 34 &             & 02 47 00.6 & +36 45 31 &    E   & 14994  45 &  7.57 & o  \\
 35 &             & 02 46 55.1 & +36 47 02 &   S0   & 14774  69 &  4.11 & o  \\
 36 &             & 02 46 43.9 & +36 47 31 &    S   & 13032  49 &  7.39 & o  \\
 37 &             & 02 46 43.5 & +36 48 32 &    E   & 13074  67 &  6.75 & o  \\
 40 & LEDA074158  & 02 46 07.6 & +36 48 18 &    S   & 14621  65 &  8.08 & o  \\
 41 &             & 02 46 10.9 & +36 47 01 &   S0   & 15170  79 &  5.20 & o  \\
 42 &             & 02 45 46.7 & +36 47 45 &    S   & 12829  85 &  5.59 & o  \\
 43 &             & 02 45 18.0 & +36 45 51 &    S   & 41230  92 &  2.94 & o  \\
 44 &             & 02 45 33.7 & +36 46 40 &   S0   & 13034 126 &  3.18 & o  \\
 45 &             & 02 45 13.1 & +36 46 28 &  S0/a  & 14356  77 &  4.97 & o  \\
 49 &             & 02 44 42.2 & +36 45 10 &   S0   & 14338  74 &  4.69 & o  \\
 52 &             & 02 47 36.2 & +36 51 04 &   S/I  & 14357  39 &  7.85 & o  \\
 54 &             & 02 47 30.1 & +36 52 03 &   S0   & 14329  30 &  8.02 & o  \\
 55 &             & 02 46 58.0 & +36 52 39 &  S0/a  & 15203  98 &  3.32 & p  \\
 57 &             & 02 46 50.1 & +36 53 21 &    S   & 15168  94 &  3.26 & p  \\
\hline
\end{tabular}
\end{table*}

\begin{table*}[htb]
\begin{tabular}{llllllll}
    \multicolumn{3}{l}{\textbf{Table 1.} continued.}\\[4pt]
\hline
\hline
Galaxy & Name & R.A. & Decl. & Type & Hel. Vel. & R & N \\
id. &     & (2000) & (2000) &    & $V {\pm {\Delta}V}$  & & \\
\hline
 60 &             & 02 46 24.1 & +36 52 35 &   S0   & 13911  86 &  6.75 & o  \\
 61 &             & 02 46 24.5 & +36 54 04 &   S0   & 14143  78 &  6.41 & o  \\
 63 &             & 02 46 20.7 & +36 55 16 &   S0   & 14058  89 &  6.70 & o  \\
 64 &             & 02 46 20.8 & +36 54 47 &   S0   & 13837  74 &  7.16 & o  \\
 65 &             & 02 46 18.5 & +36 54 52 &   S0   & 14100  54 & 11.90 & o  \\
 68 &             & 02 46 10.4 & +36 55 11 &  E/S0  & 14652  45 &  6.79 & o  \\
 69 &             & 02 46 08.9 & +36 54 22 &    E   & 15066  72 &  6.05 & o  \\
 70 &             & 02 46 07.8 & +36 54 12 &    E   & 14547  93 &  5.64 & o  \\
 71 & MCG+6-07-04 & 02 46 04.0 & +36 54 18 &   E/D  & 14563  34 &  8.54 & o  \\
    & UGC 02232   &            &           &        & 14549  39 &       & l1 \\
    & (Central galaxy)  &            &           &        & 14700     &       & l2 \\
 72 &             & 02 46 03.3 & +36 53 03 &   S0   & 16228  69 &  5.73 & o  \\
 73 &             & 02 45 54.2 & +36 49 59 &   S0   & 15988  31 & 10.53 & o  \\
 75 &  GIN 136    & 02 45 48.3 & +36 51 13 &    E   & 13620  33 &       & l1 \\
 76 &  GIN 137    & 02 45 43.8 & +36 51 17 &    E   & 13983  66 &  8.25 & o  \\
    & LEDA074141  &            &           &        & 14142  48 &       & l1 \\
 78 &             & 02 45 41.9 & +36 50 59 &    S   & 13661  49 &  4.66 & o  \\
 79 &             & 02 45 33.5 & +36 51 43 &   S0   & ~~~---    &       & o2 \\
 81 &             & 02 45 28.6 & +36 50 58 &   S0   & 14256  89 &  3.58 & o  \\
 82 &             & 02 45 22.6 & +36 50 32 &  S0/S  & ~~~---    &       & o3 \\
 83 &             & 02 45 16.5 & +36 50 45 &   S0   & ~~~---    &       & o4 \\
 84 &             & 02 45 17.7 & +36 55 05 &   S0   & 13371  71 &  5.69 & o  \\
 85 &             & 02 45 14.4 & +36 53 10 &   S0   & 14354  74 &  5.54 & o  \\
 86 &             & 02 44 50.0 & +36 55 06 &    S   & 13317  79 &  4.70 & o  \\
 89 &             & 02 47 29.2 & +36 57 44 &  S/S0  & 12550  39 &  6.74 & o  \\
 91 &             & 02 46 57.6 & +37 01 11 &   S0   & 16019  26 & 10.65 & o  \\
 92 &  GIN 138    & 02 46 50.0 & +36 58 45 &   S0   & 14798  27 & 12.98 & o  \\
    &             &            &           &        & 14825  21 &       & l1 \\
 93 &             & 02 46 29.8 & +36 58 37 &    E   & 15764  34 &  9.17 & o  \\
 94 &             & 02 46 22.7 & +36 57 38 &   S0   & 15969  34 & 12.83 & o  \\
 95 &             & 02 46 06.3 & +36 58 42 &    S   & 15012  28 & 10.74 & o  \\
 96 &             & 02 46 06.1 & +36 57 10 &    E   & 13552  37 & 10.18 & o  \\
 97 &             & 02 45 55.3 & +36 57 10 &   S0   & 13974  37 &  8.46 & o  \\
 98 &             & 02 45 50.2 & +36 57 20 &   S0   & 14861  36 &  9.81 & o  \\
100 &             & 02 45 43.8 & +36 57 26 &   S0   & 13831  89 &  5.46 & o  \\
101 &             & 02 45 44.3 & +36 56 34 &   S0   & 15196  77 &  6.17 & o  \\
102 &             & 02 45 38.3 & +36 59 04 &  S/S0  & 12722  46 &  9.30 & o  \\
103 &             & 02 45 36.1 & +36 58 19 &    S   & 16162  86 &  9.02 & o  \\
104 &             & 02 45 28.2 & +36 57 25 &   S0   & 13789  71 &  4.30 & o  \\
106 &             & 02 44 53.5 & +36 58 17 &  S0/S  & 14557  71 &  5.02 & o  \\
111 &             & 02 45 09.7 & +37 05 44 &    S   & 13749  72 &  6.96 & o  \\
\hline
\end{tabular}
\begin{flushleft}
 Notes:\\
 \textbf{l1} from Wegner \etal (1999); \textbf{l2} from Postman and Lauer 1995).\\
 \smallskip
 \textbf{1} emission line: O\textsc{ii} $=39998\kms$;\\
 \textbf{2} aligned star with a foreground galaxy ($V < 10000~\kms$);\\
 \textbf{3} dominant M~star;\\
 \textbf{4} dominant star;
\end{flushleft}
\end{table*}
\makeatletter\if@referee\renewcommand\baselinestretch{1.5}\fi\makeatother

\begin{enumerate}

\item Dressler (1980) number;

\item alternative name LEDA, UGC, GIN or MCG;

\item right ascension (J2000);

\item declination (J2000);

\item morphological type from Dressler's (1980) catalogue;

\item heliocentric radial velocity with its error in $\kms$;

\item R-value derived from Tonry \& Davis (1979);

\item telescope and notes, \textbf{o}: 1.93m OHP telescope + CARELEC
spectrograph; \textbf{p} 2.0m Pic du Midi telescope + ISARD spectrograph.\\

\end{enumerate}

Note that no velocity could be derived from objects number 17, 79, 82, 83
because of the presence of a foreground star. For galaxies already observed by
Wegner \etal (1991) and/or Postman and Lauer (1995), velocity comparison was
made separately for objects number 25, 28, 71, 76, and 92. We obtain
$<V_{o}-V_{\rm ref}> = -24\kms$, the standard deviation of the difference being
$66\kms$. These results are consistent with the errors of Table~1. The
velocities in the present study agree with those previously published within
the $2 \sigma$ level.

\section{Galaxy distribution} \label{Galaxy distribution}

We present in Figure~1 the adaptative density map of the projected
distribution of galaxies brighter than $O \leq 18^m$ obtained from the POSS I
Revised APS Catalogue\footnote{The University of Minnesota \textit{Automated
Plate Scanner}; the $O$-band is the characteristic band of the blue plates of
the First Palomar Sky Survey (Minkowski \& Abell 1963).}, in a $71 \times
90\,$arcmin$^2$ region centered on the position of the brightest cluster
member of Abell~376, E/D galaxy number~71 of Table~1. This corresponds to a
region of about $4.0 \times 5.1 h^{-2}_{70}\,$Mpc$^2$ at the distance of
Abell~376 ($z = 0.048$, see below). The central square region, of about $37
\times 44\,$arcmin$^2$,indicates the area where redshifts have been measured
(see previous Section). The positions of the brightest galaxies have been
superimposed to the density map.

\begin{figure*}[htb]
    \centering
    \includegraphics[width=12cm]{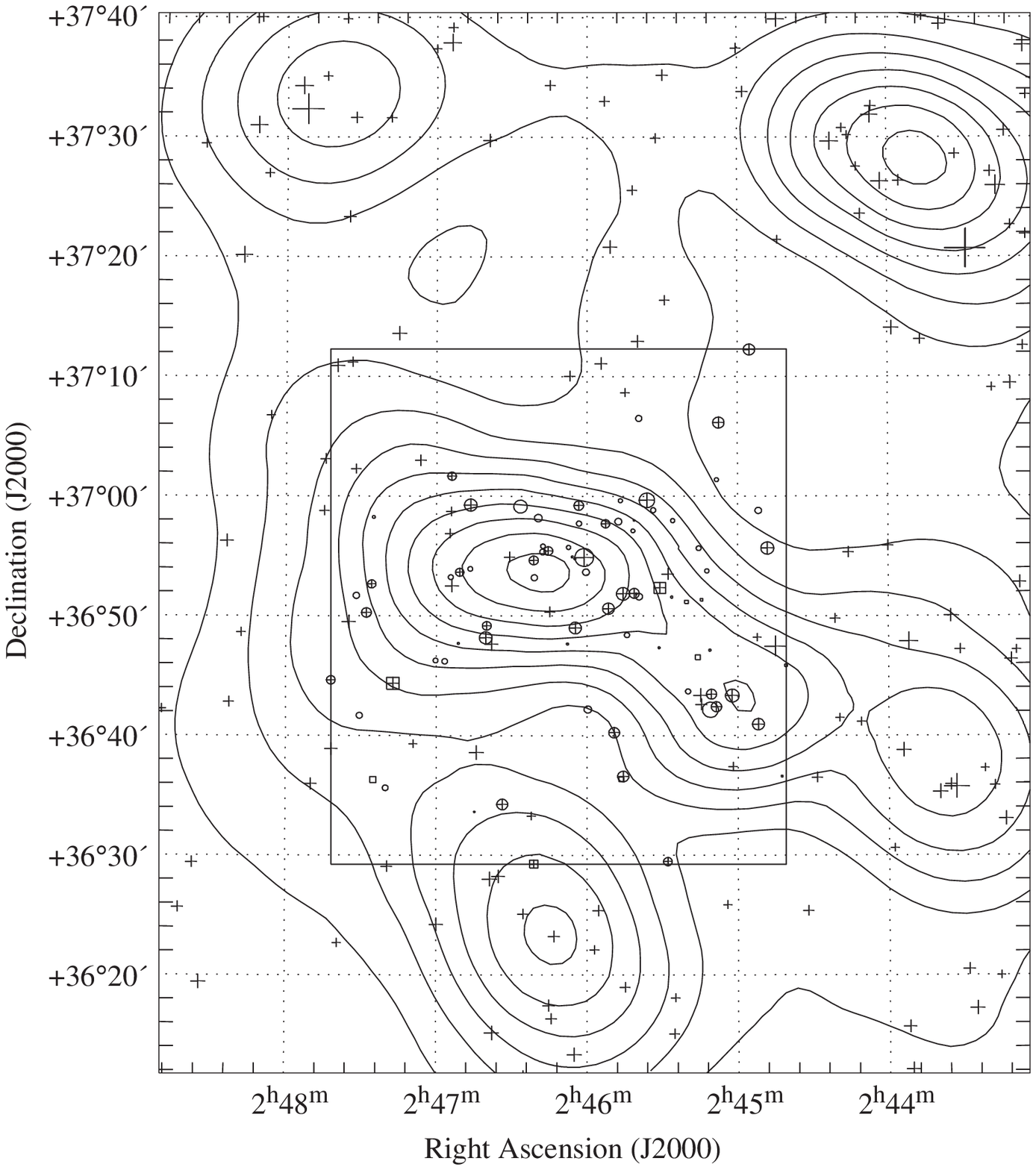}
    \caption[]{Projected density map of the 168 galaxies brighter
    than $O = 18^m$ in the field of Abell~376 together with
    the positions of the galaxies (pluses). Galaxies with
    measured redshifts are marked with circles. Back and foreground galaxies
    (see Section \ref{Velocity analysis}) are marked with squares. Larger
    symbols denote brighter galaxies.}
    \label{fig:dens_mag18}
\end{figure*}

As it can be seen from this figure, the entire field is filled with structures
among which Abell~376 stands as the central dominant one. The density map
suggests that Abell~376 has a regular galaxy distribution, flattened in the
NE-SW direction. A substructure is present at SW, which is centered on a clump
of bright galaxies around galaxy number~28, classified as E by Dressler
(1980). As discussed in the next section, this clump may be considered as
kinematically belonging to Abell~376 cluster. This figure also indicates that the
cluster radial extension may attain several Mpc.

However since Abell~376 lies close to the galactic plane ($b \approx
-20.6^{\circ}$), before trying to discuss its apparent projected galaxy
distribution, we should be confident that the galactic extinction is not
significantly affecting the magnitudes and counts. Figure~2 shows the map of
galactic absorption estimated from Schlegel \etal (1998), with a resolution of
2.5~arcmin. The colour excess E($B-V$) was transformed to the $O$-magnitudes
using the appropriate transmission curve from Larsen (1996), given in the APS
site\footnote{\texttt{http://aps.umn.edu/Docs/photometry.aps.html}} and the
Table~6 from Schlegel \etal (1998). As it can be seen, although this map bears
a vague resemblance with density the distribution of galaxies given in
Figure~1, the absorption $A_O$ never attains values higher than $\sim 0.4$ and
thus it should not significantly affect the apparent distribution of galaxies.

\begin{figure*}[htb]
  \centering
  \includegraphics[width=12cm]{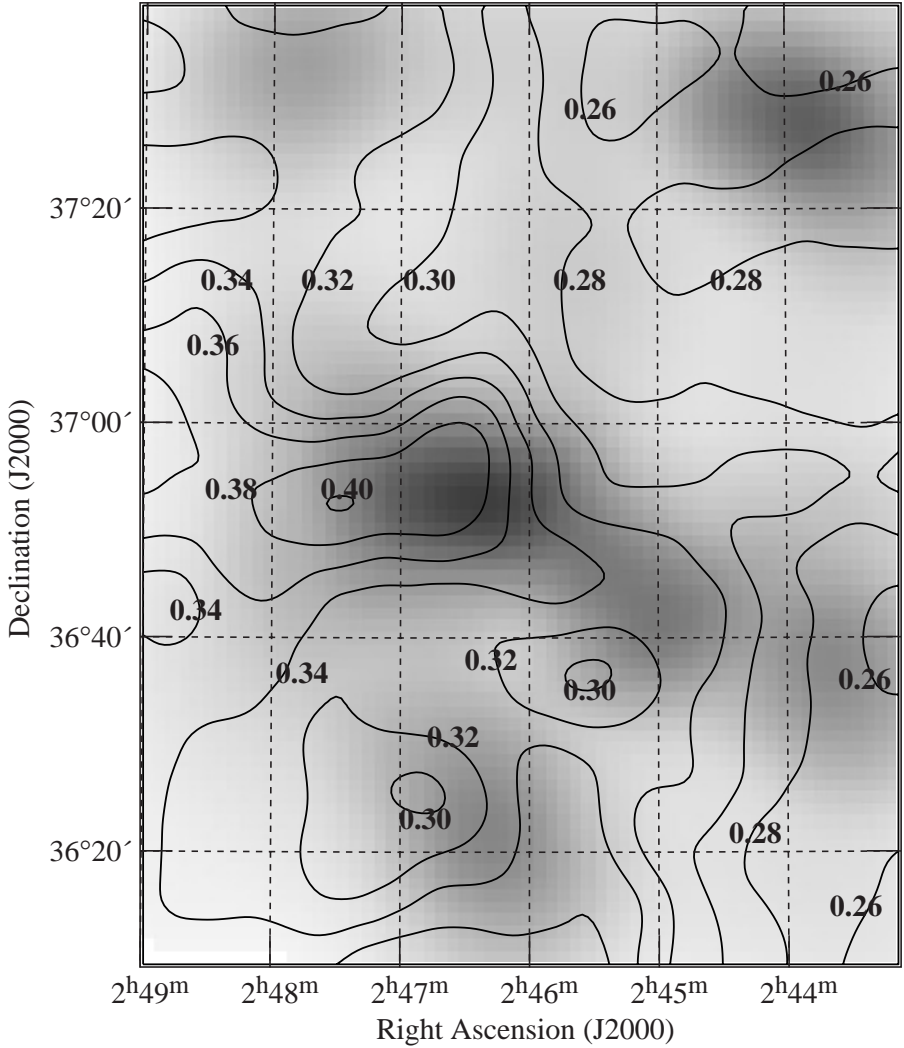}
  \caption[]{Contours of constant galactic absorption, $A_O$ (labelled),
   superimposed to the surface density image of galaxies brighter
   than $O = 18^m$.}
  \label{fig:absorption}
\end{figure*}

Considering only galaxies brighter than $O = 18^m$, the central peak of
density is displaced by $\sim 4.7\,$arcmin~E, relative to the central E/D
galaxy. However there is no such a displacement when considering galaxies
brighter than $O = 19.5^m$ (see also Figure~4). For $O = 18^m$, central
densities amount to $\simeq {2.8 \cdot 10^2}\,\mbox{deg}^{-2}\, (\simeq {60}
h^2_{70}\,$Mpc$^{-2})$.

There is no redshift information about the dense NW condensation seen in
Figure~1 since redshifts measurements are restricted in a square region around
Abell~376. However a hint about its relative distance to Abell~376 may be
given by comparing the apparent magnitude distribution of galaxies in
Abell~376 with that of the NW condensation. This comparison is made in
Figure~3, where we also show the magnitude distribution of galaxies with
measured redshifts. As it can be seen, both histograms, for Abell~376 and for
the NW condensation, show almost the same slope, suggesting that both are
about the same distance and probably belong to the same superstructure.

\begin{figure}[htb]
    \centering
    \includegraphics[width=8.6cm]{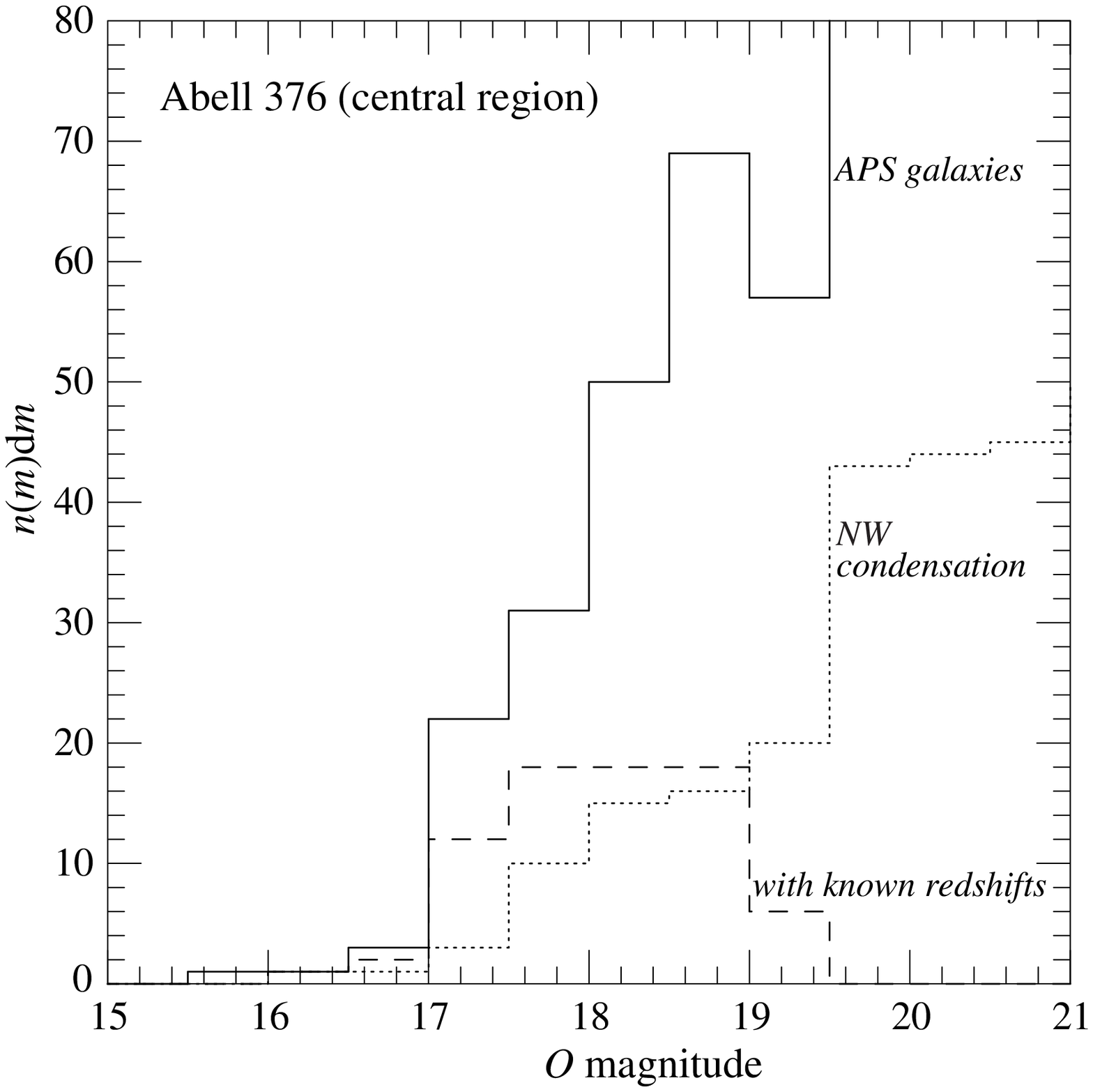}
    \caption[]{$O$ magnitudes histograms: \textit{continuous line}:
    galaxies inside the square region of measured redshifts; the
    \textit{dashed line} histogram is for galaxies with measured
    redshifts (see Figure~1). \textit{Dotted lines}: galaxies belonging
    to the NW condensation.}
    \label{fig:fdl}
\end{figure}

The histogram in Figure~3 also indicates that the sample of measured radial
velocities in the central region of Abell 376 is $\sim 60$\% complete at
$O_{\rm lim} = 18^m$. This is indeed a very shallow magnitude limit, given the
distance of this cluster. Figure~4 shows a comparison of the surface density
map of projected galaxies brighter than $19.5^m$ together with the redshift
sample of galaxies discussed above, which, as discussed in the next Section,
comprises galaxies kinematically members of the cluster. As it can be seen,
excepted in the SE direction, the galaxies with measured redshift sample well
the projected galaxy distribution. Therefore, despite the incompleteness of
the redshift sample, it will be possible to discuss the line of sight (l.o.s.)
velocity distribution at least in a qualitative way.

\begin{figure}[htb]
    \centering
   \includegraphics[width=8.6cm]{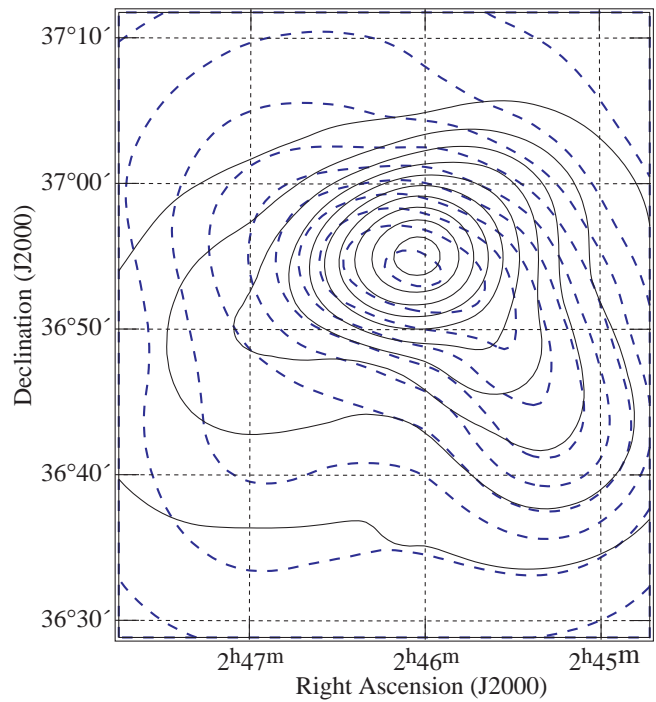}
   \caption[]{Isopleths of the projected distribution of galaxies with
   $12000 \kms <  V < 17000 \kms$ (dashed lines), together with the
   corresponding ones for APS galaxies brighter than
   $19.5^m$ (continuous lines).}
   \label{fig:dens-x-vel}
\end{figure}

\section{Velocity analysis}\label{Velocity analysis}

Including previous measurements (Capelato \etal 1991, Postman and Lauer
(1995), Wegner \etal (1999), there is a total of 73 measured velocities for
Abell~376. Using this sample we will discuss the cluster velocity distribution,
looking for non-equilibrium effects. The wedge diagrams of galaxies in right
ascension and declination are displayed in Figure~5. Both show the cluster
and all galaxies collected from the NED database in a radius of 300 arcmin
around the cluster centre (476 objects). One can see that Abell~376 does not show
any connection or extension with other evident structure.

\begin{figure}[htb]
   \centering
   \includegraphics[width=8.6cm]{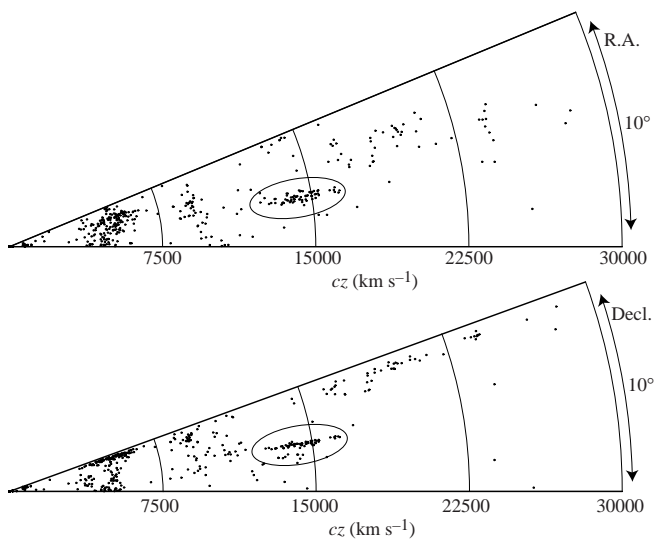} 
\caption[]{Wedge velocity diagram in right ascension (\textit{top}), and
declination (\textit{bottom}) for the galaxies in the Abell~376 field with
measured radial velocities smaller than $V < 30000\kms$. Abell~376 is situated
at $cz \sim 14500\kms$ (thin ellipsis) with a classical elongated shape,
surrounded by 403 galaxies with known redshifts within a radius of
300~arcmin from the cluster center.}
\end{figure}

We used the ROSTAT package (Beers \etal 1990; Bird \& Beers 1993) to analyze
the distribution of the 63~galaxies in the selected velocity range $12550\kms
< V < 16250\kms$ (for galaxies with more than one value of radial velocity in
Table~1, a mean value was assumed). In order to quantify the central location
and scale of the velocity distribution for Abell~376, we have used the
resistant and robust bi-weight estimators ($C_{BI}$ and $S_{BI}$,
respectively) recommended by Beers \etal 1990). For this sample of velocities,
we obtain $C_{BI} = 14329^{+144}_{-168}\kms$ and $S_{BI} =
852^{+120}_{-76}\kms$. Estimates of these quantities obtained with alternative
estimators show similar values. The one-sigma errors in these quantities are
calculated in ROSTAT by bootstrap re-sampling of 1000~subsamples of the
velocity data. For comparison, the radial velocity of the E/D galaxy located
at the centre of the main cluster is $14563 \pm 34\kms$, near that of the
whole cluster, as should be expected if this is the dominant cluster galaxy.

If we consider the morphological types, the mean velocities and corrected
velocity dispersions are: $\overline{V} = 14468\kms$,
$\sigma = 761^{+131}_{-93}\kms$ for E~$+$~S0 galaxies (48~objects) and
$\overline{V}= 14059\kms$, $\sigma = 911^{+176}_{-115}\kms$ for
S~$+$~I galaxies (22~objects). As what is observed in most clusters, the
velocity dispersion of the late-type population tends to be larger than
that of the early-type population (Sodr\'e \etal 1989, Stein 1997, Carlberg
\etal 1997, Adami \etal 1998). Figure~6 presents the radial velocity
distribution of the cluster galaxies as well as a Gaussian curve with the
same mean velocity and dispersion observed for these galaxies between 12000
and $17000\kms$. Note that the velocity distribution obtained with a larger
set of velocities does not confirm the relative depletion of galaxies with
$V \simeq 14000\kms$ proposed by Capelato \etal (1991).

\begin{figure}[htb]
\centering
\includegraphics[width=8.6cm]{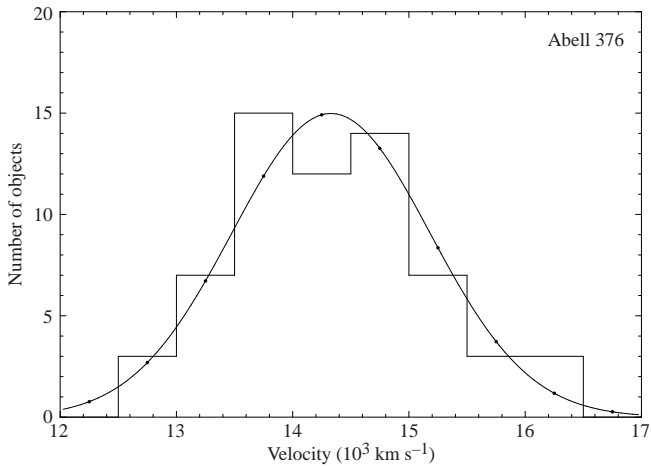}
\caption[]{The radial velocity distribution for the Abell~376 sample of
galaxies between 12000 and $17000\kms$. The continuous curve shows the
Gaussian distribution corresponding to the mean velocity and velocity
dispersion quoted in the text (normalized to the sample size and range).}
\end{figure}

\section{X-ray emission and gas distribution}\label{X-ray emission}

Abell 376 is an average X-ray source, first detected by the \textit{Uhuru}
satellite (Kellogg \etal 1973). It was then observed by \textit{Einstein} IPC,
\textit{Exosat}, ROSAT HRI and PSPC (part of the ROSAT
all-sky survey), and ASCA. Abell 376 is included in the XBACs catalogue (X-ray
Brightest Abell type Cluster, Ebeling \etal 1996).

\begin{figure*}[htb]
    \centering
    \includegraphics[width=18cm]{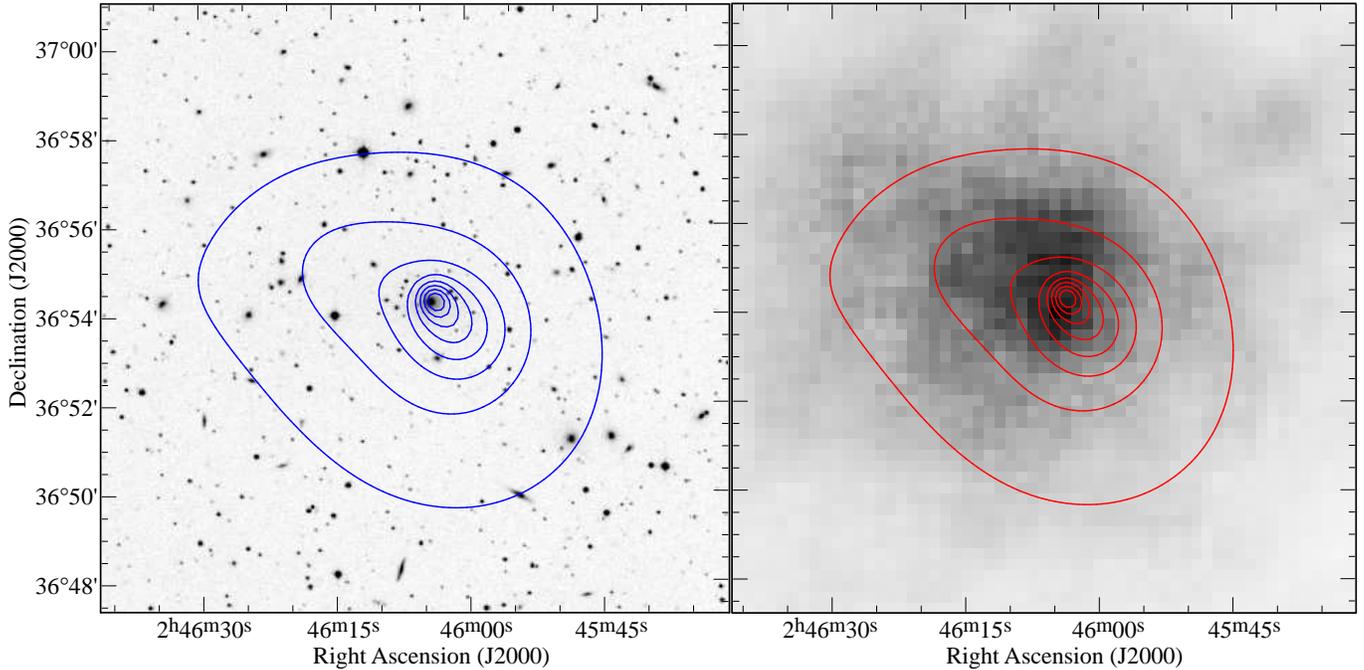}
    \caption{Left: Red DSS image overlayed with ROSAT HRI isocontours.
    Right: ASCA GIS broad band smoothed image superposed with HRI isocontours.}
    \label{fig:A376_dss_red_ASCA_HRI}
\end{figure*}

The physical properties of Abell 376 intracluster medium (ICM) were first
obtained with \textit{Exosat} data in the [0.1--20.0]~keV band (Edge \&
Stewart 1991). Using the RS thermal plasma model (Raymond \& Smith 1977) to
fit the \textit{Exosat} X-ray spectrum, they derived a mean temperature for
the ICM, $kT = 5.1^{+3.2}_{-1.9}\,$keV (90\% confidence) and an X-ray flux in
the [2.0--10.0]~keV band, $f_{\rm X} = 1.01 \times
10^{-11}$erg~cm$^{-2}$s$^{-1}$, which corresponds to a bolometric luminosity
of $L_{\rm X} = 1.19 \, h_{70}^{-2} \times 10^{44}$erg~s$^{-1}$. The
deprojection analysis of the \textit{Exosat} image suggested a mild cooling
flow, with a mass flow rate of $\dot{M} \le 48 M_{\odot}$yr$^{-1}$ and a
cooling radius $R_{\rm cool} (68\pm 26) h_{70}^{-1}\,$kpc.

The ROSAT PSPC observation of Abell 376 was analyzed as part of the XBACs
catalogue (Ebeling \etal 1996). The temperature was fixed using the EXOSAT
value (see above) and the derived flux (assuming a fixed metallicity of $0.3
Z_{\odot}$) was $f_{\rm X} = 1.27 \times 10^{-11}$erg~cm$^{-2}$s$^{-1}$ in the
[2.0--10.0]~keV band.

White \etal (1997) analyzed the \textit{Einstein} IPC X-ray image of Abell 376
with a deprojection technique. In order to be applied, they used the ICM
temperature given by Edge \& Stewart (1991) as an input to the deprojection.
The resultant temperature was thus $T_{\rm X} = 5.7^{+0.4}_{-1.5}\,$keV and
the bolometric luminosity, $L_{\rm X} = 8.42 h_{70}^{-2} \times
10^{43}$erg~s$^{-1}$. They also derived a weak cooling-flow, with a mass
deposit rate of $\dot{M} = 42^{+42}_{-14} M_{\odot}$yr$^{-1}$ inside a cooling
radius of $R_{\rm cool} (60^{+89}_{-30}) h_{70}^{-1}\,$kpc, well in agreement
with the previous result based on \textit{Exosat} data.

The imaging analysis also yields the gas and dynamical masses. Within the
outer radius, where the observation was meaningful ($R_{\rm out} = 239
h_{70}^{-1}\,$kpc), White \etal (1997) estimated $M_{\rm gas} = (5.1 \pm 1.0)
\times 10^{12} M_{\odot}$ and $M_{\rm dyn} = 175 \times 10^{12} M_{\odot}$.

From the HEASARC Online Service, we have downloaded the ROSAT HRI and
ASCA GIS data files. As far as we know, these are still unpublished data. With
the HRI data, we have created an exposure-map, vignetting corrected broad-band
image following the ``cookbook'' from Snowden \& Kuntz (1998); this image was
then adaptatively smoothed.

The ASCA GIS spectrum and image were extracted and cleaned with XSELECT 1.4b
following the ``ASCA data reduction
guide''\footnote{\texttt{http://heasarc.gsfc.nasa.gov/docs/asca/abc/abc.html}}.
We have used the GIS background files provided by the ASCA guest observer
facility (GOF). For the spectroscopy, the GIS data was selected in a circle of
8.1~arcmin centered in the BCG and in the [0.2--10.5]~keV energy band, then
rebinned with a minimum of 30 counts per energy bin.

In Fig.~\ref{fig:A376_dss_red_ASCA_HRI} we show both an optical DSS and
ASCA X-ray image superposed with the smoothed HRI image isocontours. The
X-ray centre as determined by the HRI coincides with the central elliptical
galaxy of Abell 376. There is a good agreement between HRI and GIS images,
although there seem to be a small offset between them. Such an offset is
actually comparable to the known uncertainty of the ASCA aspect solution.

\begin{figure}[htb]
    \centering
   \includegraphics[width=8.6cm]{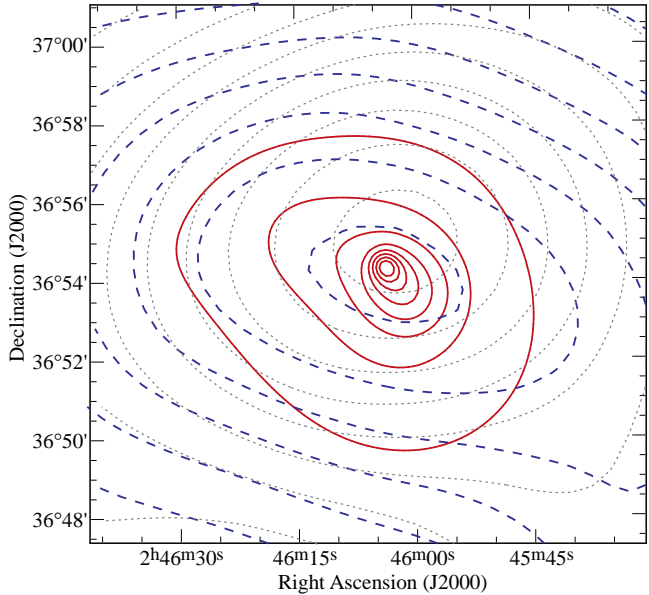}
   \caption[]{ROSAT HRI isocontours (continuous lines) as in 
   Fig.~\ref{fig:A376_dss_red_ASCA_HRI}
   superposed to the isopleths of the projected galaxy distribution (dashed
   lines, galaxies within $12000 \kms < V < 17000 \kms$; thin dotted lines,
   APS galaxies brighter than $19.5^m$) as in Fig.~\ref{fig:dens-x-vel}.}
   \label{fig:A376_HRI_Isopleta}
\end{figure}

The X-ray images, ROSAT HRI and ASCA GIS, probe only the central $\sim 
15\,$arcmin region and they show no evident sign of substructure. The X-ray 
isocontours are not spherical or elliptical and are elongated roughly in a 
Northeast--Southwest direction. Comparison between the X-ray emission and the 
cluster galaxy distribution (Fig.~\ref{fig:A376_HRI_Isopleta}) shows a good 
qualitative agreement between both gas and galaxy spatial distribution, when 
considering only objects within $12000 \kms < V < 17000 \kms$. This agreement 
may extend further on, since the orientation of the external isopleths seen in 
Fig.~\ref{fig:dens-x-vel} are close to the inner contours. When we 
take into account all APS galaxies brighter than $19.5^{m}$, there is an 
offset between X-ray emission and galaxy distribution close to the 
North-South direction.

\begin{figure}[htb]
    \centering
    \includegraphics[width=8.6cm]{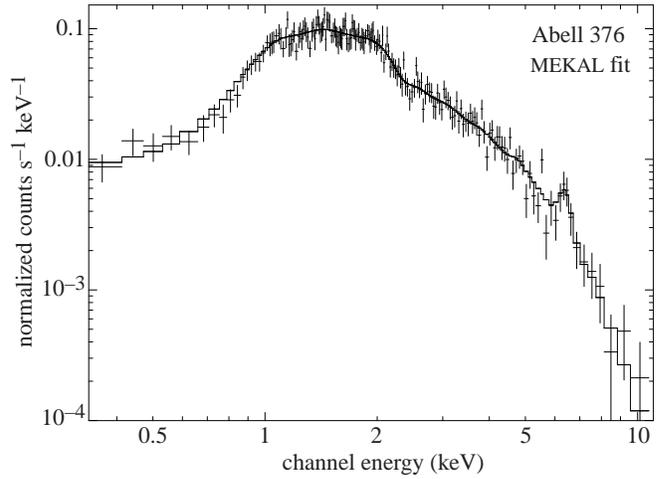}
    \caption[]{ASCA GIS spectrum showing the MEKAL plasma model best fit. The
    fit shown here have a fixed $N_{\rm H} = 6.76 \times 10^{20}$cm$^{-2}$.}
    \label{fig:gis3Abell376mekal}
\end{figure}

The ASCA X-ray spectrum was fitted with a single temperature, absorbed MEKAL
model (Kaastra \& Mewe 1993; Liedahl \etal 1995) using XSPEC~11. We have
tried two fits fixing the hydrogen column density to the galactic value,
$N_{\rm H} = 6.76 \times 10^{20}$cm$^{-2}$ (Dickey \& Lockman, 1990), and
allowing it to vary as a free parameter. The results are summarized in
Table~\ref{tbl:fitting} and shown in Fig.~\ref{fig:gis3Abell376mekal}

\begin{table}[htb]
 \centering
 \caption{Fitting results of the ASCA GIS spectrum. Errors bars are 90\% 
 confidence level. The last column gives the $\chi^{2}$ over the number of 
 degrees of freedom (d.o.f.).}
 \begin{tabular}{c c c c}
     \hline
   $N_{\rm H}$        & $kT$  & $Z/Z_{\odot}$ & $\chi^{2}$/d.o.f. \\
   $10^{20}$cm$^{-2}$ & (kev) & & \\
     \hline
    6.76          & $4.3 \pm 0.4$ & $0.32 \pm 0.08$ & 130.0/173 \\
   $14.7 \pm 3.5$ & $3.6 \pm 0.4$ & $0.43 \pm 0.10$ & 122.4/172 \\
     \hline
 \end{tabular}
 \label{tbl:fitting}
\end{table}

The mean temperature obtained depends whether $N_{\rm H}$ is fixed or not (a
well known anti-correlation between spectral fitted temperature and $N_{\rm
H}$), but are within error bars. The temperatures we got are slightly cooler
than the ones obtained with ROSAT and \textit{Exosat}. The same happens with 
the metal abundance; when $N_{\rm H}$ is fixed the obtained metallicity is lower.

The derived flux and luminosity are about the same with the two fits in
Table~\ref{tbl:fitting}: in the [2.0--10.0 keV] band we have $f_{X} = (0.97
\pm 0.09)\, 10^{-11}\,$erg~s$^{-1}$cm$^{-2}$ and $L_{X} = (5.4 \pm
0.5)\,h_{70}^{-1}\, 10^{43} \,$erg~s$^{-1}$.

Using the scaling relations obtained by Xue \& Wu (2000) we have $\sigma = 
520^{+220}_{-120}\kms$ (with the $L_{X}$--$\sigma$ relation), and $\sigma = 
760^{+85}_{-80}\kms$ (with the $T_{X}$--$\sigma$ relation). Here we have 
adopted the mean temperature of the two fits, $kT = 4.0\,$keV. The value of 
$N_{\rm H}$ is, however, ill constrained and depends on the few spectral 
points below $\sim 1\,$keV and a better value for the temperature may be the 
one from the first fit, $kT = 4.3\,$keV. In this case, $\sigma= 
798^{+75}_{-70}\kms$.

\section{Morphological and spectral classification}\label{spectral classification}

It has been shown that it is possible to classify galaxies from their spectra,
and that this spectral classification correlates well with morphological types
(e.g., Folkes \etal 1996, Sodr\'e \& Cuevas 1997, Madgwick \etal 2002). Since
for Abell~376 we have the morphological classification done by Dressler
(1980), it is interesting to verify how his Hubble types compare with spectral
types obtained from galaxy spectra.

The basis of spectral classification is that normal galaxy spectra
form a well defined sequence in the spectral space spanned by the
$M-$dimensional vectors that contain the spectra, each vector
being the flux of a galaxy (or a scaled version of it) sampled at
$M$ wavelengths. We define the spectral type of a galaxy by its
position along the spectral sequence, identified through a
principal component analysis (PCA) of the galaxy spectra (see
Sodr\'e \& Cuevas 1997 and Sodr\'e \& Stasi\'nska 1999 for more
details).

Here, the analysis has been done using 54 spectra of cluster
galaxies with high signal-to-noise ratios. Initially the spectra
were shifted to the rest frame and re-sampled in the wavelength
interval from 3784~\AA\ to 6100~\AA, in equal bins of 2~\AA.
After, we removed from the analysis 8 regions of $\sim 20$\AA\ 
each centered at wavelengths that may contain emission lines,
because the inclusion of emission lines in the analysis increases
the dispersion of the spectral sequence. The spectra were then
normalized to the same mean flux ($\sum_\lambda f_\lambda = 1$)
and the mean spectrum was computed and removed from the galaxy
spectra. PCA was then used to obtain the principal components.

Figure~9 shows the projection of the 54 spectra on to the plane
defined by the first two principal components. The first principal
component contains 39\% of the sample variance, whereas the second
component carries 7\% of the variance. We identify the spectral
type ST of a galaxy with the value of its first principal
component. In Fig.~9 different symbols correspond to different
morphological types: filled circles, open squares, open triangles,
and crosses represent E, S0, S0/S, and S galaxies, respectively.
The mean values of ST for each of these 4 classes are $-0.78
\times 10^{-3}$, $-0.22 \times 10^{-3}$, $0.24 \times 10^{-3}$,
and $0.98 \times 10^{-3}$, respectively. This result confirms that
the spectral sequence is closed related to the morphological
sequence, since the ranking of galaxies of different morphological
types along the first principal component is analogous to that
found in the Hubble sequence, despite the strong overlap in the
values of ST between different classes. Note that although the
values of the principal components are dependent on the data and
its pre-processing, the ranking of galaxies along the first
principal component is in large extent independent of the
pre-processing adopted.

\begin{figure}[htb]
    \centering
\includegraphics[width=8.6cm]{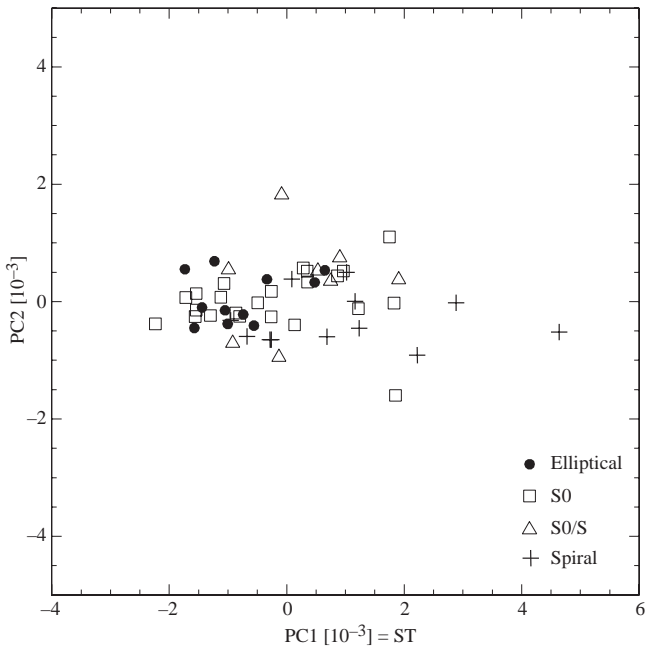}
\caption[]{Projection of the spectra of 54 cluster galaxies on to
the plane defined by the first two principal components. The
spectral sequence follows the first component and a spectral type,
ST, is attributed to each galaxy by its value of this component.
Filled circles, open squares, open triangles, and crosses
represent E, S0, S0/S, and S galaxies, respectively.}
\end{figure}

\section{Cluster virial mass}\label{mass}

We have estimated the cluster mass inside its virial radius with the virial
mass estimator (VME), which gives less biased mass estimates when the system
is not completely sampled (e.g., Aceves \& Perea 1999). The virial radius of
Abell~376 was estimated through an iterative procedure with the relation $R_V
\simeq 0.003 \sigma_p\, h_{70}^{-1}\,$Mpc (Girardi \etal 1998), where
$\sigma_p$ is the projected velocity dispersion.

Initially, we estimated the virial radius from the 62~galaxies with radial
velocities between 12000~km~s$^{-1}$ and 17000~km~s$^{-1}$. Then, considering
only galaxies within the virial radius, $\sigma_p$ was re-evaluated and a new
virial radius was computed. New iterations do not changed the value of the
virial radius, $R_V \simeq 2.58 h_{70}^{-1}\,$Mpc ($\sim 42\,$arcmin), which
is larger than the 28 arcmin ($1.6~h_{70}^{-1}$ Mpc) central region where
radial velocities were measured.

The resulting mass within $0.7 R_{V}$ is $(1.05 \pm 0.27) \times 10^{15}
h_{70}^{-1}M_\odot$, where the error is a $1 \sigma$ standard deviation
computed with the bootstrap method. Note that the VME assumes that the system
is virialized, but substructures or large-scale flows tend to increase the
velocity dispersion of the cluster galaxies, leading to an overestimation of
the cluster mass. In a radius of 28 arcmin, the only feature visible is the 
SW substructure, that is not even detected in the velocity-space (see 
Sect.~\ref{Velocity analysis}),

\section{Conclusion}\label{Conclusion}

In this paper, we have reported new velocity measurements on the Abell~376
cluster of galaxies and an X-ray analysis of previously unpublished ASCA data.
This allowed us to define almost complete samples of galaxies in the central
part of the cluster, and to discuss the dynamical status of this cluster. The
study of the galaxy projected positions reveals a regular distribution,
centered on an E/D dominant galaxy. The analysis with the adaptive kernel
density map indicates the presence of a substructure SW of the cluster main
galaxy concentration which is not confirmed by the velocity field, implying
that it may be a fortuitous condensation of galaxies bound to Abell~376. Our
dynamical analysis give for Abell~376 a virial mass about $(1.1\pm 0.3) \times
10^{15}~h_{70}^{-1}M_\odot$ within 70\% of the virial radius of the cluster.
The velocity dispersion derived from the X-ray analysis is in agreement with
the galaxy velocity distribution. We conclude that Abell~376 is a \textit{bona
fide} relaxed cluster inside $\sim 900h_{70}^{-1}\,$kpc or $\sim 40$\% of the
virial radius, fairly relaxed inside $0.7 R_{V}$ with only a minor 
substructure, and clumpier as we reach the virial radius, where possible 
substructures are detected in the galaxy distribution.

\acknowledgements{We thank the OHP and Pic du Midi staff for their assistance
during the observations. HVC, LSJ and GBLN thank the financial support
provided by FAPESP and CNPq. DP acknowledges the France-Brazil PICS-1080 and
IAG/USP for its hospitality.}

\end{document}